\documentclass{PoS}

\title{Implications of the top (and Higgs) mass for vacuum stability}

\ShortTitle{Implications of $M_t$ (and $M_h$) for vacuum stability}

\author{Jos\'e R. Espinosa
\\
ICREA, Instituci\'o Catalana de Recerca i Estudis Avan\c{c}ats, Barcelona, Spain\\
Institut de F\'isica d'Altes Energies (IFAE), The Barcelona Institute of Science and Technology (BIST), Campus UAB, E-08193, Bellaterra (Barcelona), Spain\\
        E-mail: \email{espinosa@ifae.es}}

\abstract{The discovery of the Higgs boson by the LHC and the measurement of its mass at around 125 GeV, taken together with the absence of signals of physics beyond the standard model, make it possible that we might live in a metastable electroweak vacuum. 
Intriguingly, we seem to be very close to the boundary of stability and this near-criticality makes our vacuum extremely long-lived. In this talk I describe the state-of-the-art calculation leading to these results, explaining what are the ingredients and assumptions that enter in it, with special emphasis on the role of the top mass. I also discuss possible implications of this metastability for physics beyond the standard model and comment on the possible impact of physics at the Planck scale on near-criticality.}

\FullConference{8th International Workshop on Top Quark Physics, TOP2015\\
		14-18 September, 2015\\
		Ischia, Italy}

\begin{document}

\section{The metastability of the electroweak vacuum after the first LHC run}

\begin{figure}[b]
$$\includegraphics[width=0.47\textwidth]{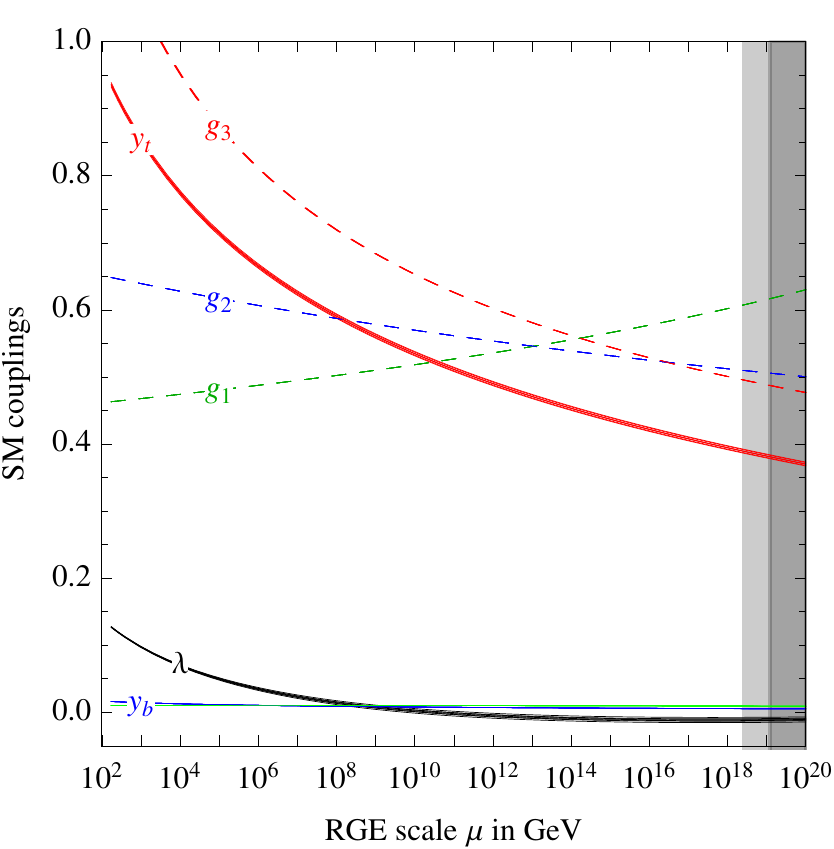}    \qquad
\includegraphics[width=0.49\textwidth]{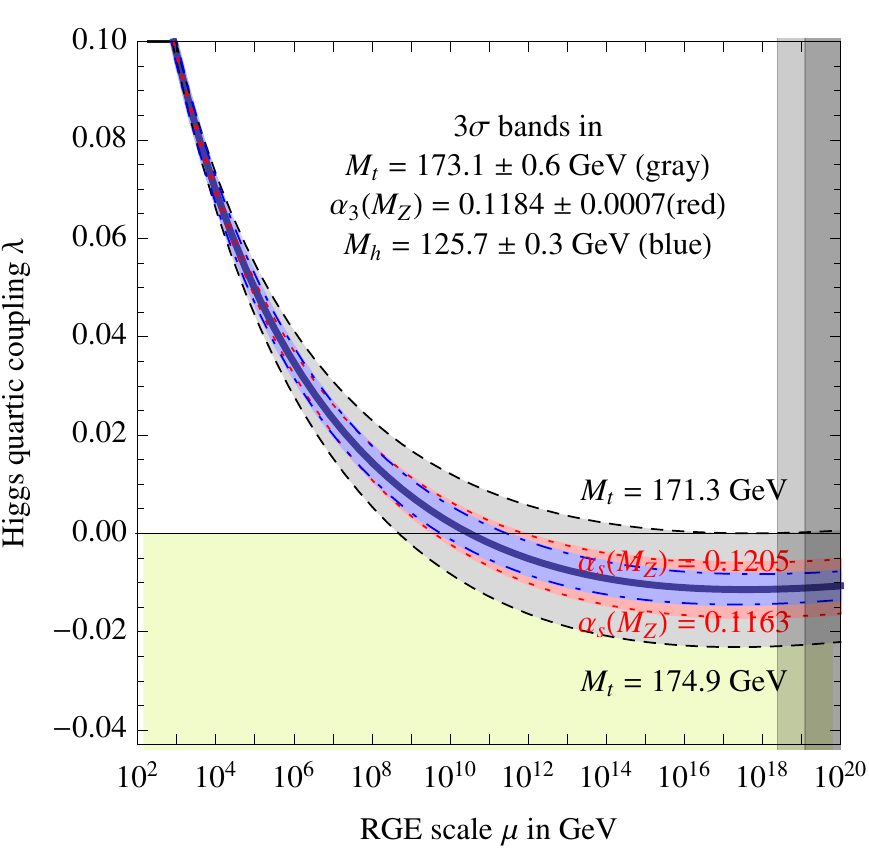}$$
\begin{center}
\caption{\label{fig:run}\emph{Left: Extrapolation of SM couplings from the Fermi scale to $M_{Pl}$. Right: Zoom-in on the  evolution of the Higgs quartic coupling, $\lambda(\mu)$, for $M_h=125.7$\,GeV. The 3$\sigma$ uncertainties in $M_t$, $\alpha_s$ and $M_h$ are shown by the colored intervals as indicated. (Taken from Ref.~\cite{us}).}}
 \end{center}
\end{figure}

In the first LHC run we have learned that the Higgs boson exists; it is light, with mass $M_h\simeq 125$\,GeV~\cite{higgsdiscovery}; and it has SM-like couplings (still with room for significant deviations). Moreover, no trace of BSM physics has showed up, leading to bounds on the mass scale $\Lambda$ of new physics in the TeV range for the main BSM scenarios, supersymmetric or not. For those of us willing to hold on to the naturalness paradigm, the hierarchy problem affecting electroweak symmetry breaking implies that new physics should be around the corner, likely on the reach of the second LHC run. 
However, it is also possible that naturalness has mislead us and we are just seeing evidence that the SM is all there is up to very high energy scales, possibly up to $\Lambda\sim M_P$.
Figure~\ref{fig:run} (left plot) shows how the most relevant SM couplings evolve when extrapolated to very high scales~\cite{us}.
It was not guaranteed but the theory stays weakly coupled up to $M_P$ but it does. We see the three  gauge couplings almost unifying at $\mu \sim 10^{14}$ GeV. The top Yukawa coupling decreases at high energy (due to $\alpha_s$ effects) and eventually becomes smaller than all gauge couplings. The Higgs quartic coupling evolves in a very interesting way: it is small at the EW scale, $\lambda(M_t)\sim 1/8$, as the Higgs boson is light, and it decreases when run to higher scales. The zoomed-in right plot in Fig.~\ref{fig:run} shows $\lambda$ becoming negative at $\mu\sim 10^{10}$\,GeV.

The steep slope of $\lambda(\mu)$ is caused by one-loop top  corrections, that give the dominant contribution to $\beta_\lambda=d\lambda/d\log\mu$, which dictates the evolution of $\lambda$ with scale. One has $\beta_\lambda =-6y_t^4/(16\pi^2)+...$ 
where $y_t$ is the sizable top Yukawa coupling. This dependence of $\beta_\lambda$ on the fourth power of $y_t$ explains the crucial sensitivity of the running of $\lambda$ on the top quark mass $M_t$, illustrated by the gray band in Fig.~\ref{fig:run} (right), corresponding to a $3\sigma$ interval of $M_t$ around is central value. The larger (smaller) $M_t$ is, the steeper (softer) the slope of $\lambda(\mu)$. The running of $\lambda$ has a smaller sensitivity to $\alpha_s$, which affects $\beta_\lambda$ indirectly through its effect on the running $y_t(\mu)$.
The effect is illustrated in Fig.~\ref{fig:run} (right) by the thinner $3\sigma$ red band, with larger (lower) $\alpha_s$ leading to softer (steeper) running. The thinnest blue band, corresponds to $3\sigma$ changes of $M_h$. We also see that $\lambda$ flattens out after becoming negative: in that range of large scales the gauge couplings are comparable in size to $y_t$ (Fig.~\ref{fig:run}, left) and their positive contribution to $\beta_\lambda$ balances the top one, leading to   $\beta_\lambda\simeq 0$. 

The trouble with $\lambda<0$ is that it leads to an unstable Higgs potential\footnote{This instability problem due to a heavy fermion coupled to a light scalar was known for a long time~\cite{oldies} and was studied since then in the SM with refined degree of precision~\cite{stability}, especially before the Higgs discovery~\cite{us0,lindner,shap,us,Buttazzo} when it was already clear that $M_h$ would be in a mass region critical for the stability of the potential. }: at high field values this potential is dominated by the quartic term, and a good approximation to the full potential at such large field values $h$ requires that the couplings should be evaluated at a renormalization scale $\mu\sim h$. Therefore,  $V(h\gg M_t) \simeq (1/4)\lambda(\mu=h) h^4$, and for $\lambda(h)<0$ the potential is deeper than the EW vacuum, which is no longer the true vacuum. We should therefore worry about the lifetime of our vacuum against decay through quantum tunneling down to larger field values that probe the instability region.

The unstable EW vacuum can decay by nucleation of bubbles that probe the instability region and are large enough to grow, eating the whole of space. The probability of such EW vacuum decay is given by the decay-rate per unit time and unit volume~\cite{tunnel} $\sim h_t^4 \exp(-S_4)$, with $h_t$ the Higgs field value beyond the region of instability to which the tunneling occurs (the only relevant mass/energy scale), and with $S_4$ 
the action of the 4D Euclidean bounce solution for tunneling that interpolates between the EW phase and $h_t$.
A simple analytical approximation obtained for a negative-quartic potential $V\simeq -|\lambda(h)|h^4/4$, gives $S_4\simeq -8\pi^2/(3|\lambda(h_t)|)$ and captures the parametrics of the main effect.\footnote{The tunneling rate in \cite{us,Buttazzo}  is calculated beyond tree level and includes the effects of fluctuations around the bounce~\cite{rateloop}. Gravitational corrections, which have a negligible impact on the rate, were also included as in Ref.~\cite{rategrav}.} The logarithmic dependence of $\lambda(h)$ on $h$ breaks the scale invariance of the classical quartic potential so that the tunneling takes place preferentially through bubbles at the scale $h_t$ at which $\lambda(h)$ reaches its minimum [that is, $\beta_\lambda(h_t)=0$]. The decay rate is then $dp/(dV \ dt)\sim h_t^4 \mathrm{exp}[-2600/(|\lambda|/0.01)]$. This tiny number has to be multiplied by the huge 4D spacetime volume of our past lightcone, which is basically given by the fourth power of the age of the Universe $\sim \tau_U^4\sim (e^{140}/M_{Pl})^4$. We find that the exponential suppression of the decay rate [for the typical $\lambda(h_t)\sim -0.01$] wins over the volume factor, resulting in a decay probability that is extremely suppressed: $p\ll 1$. In other words, the EW vacuum lifetime, $\tau_{EW}$, is extremely long, exceeding by a huge factor the age of the Universe. From the large value of the vacuum lifetime we can already conclude that the instability of the SM potential does not require the existence of new physics that stabilizes the potential: the instability does not represent a fatal blow to the consistency of the model. This reassuring conclusion could have been different if $M_h$ were smaller. In such cases a lower $\lambda(\mu)$ could enter the dangerous region $\lambda(\mu)< -0.05$ which corresponds to a vacuum lifetime $\tau_{EW}<\tau_U$ (the really dangerous instability region).

\section{Near-criticality and implications}

\begin{figure}
$$ 
\includegraphics[width=0.44\textwidth,height=0.44\textwidth]{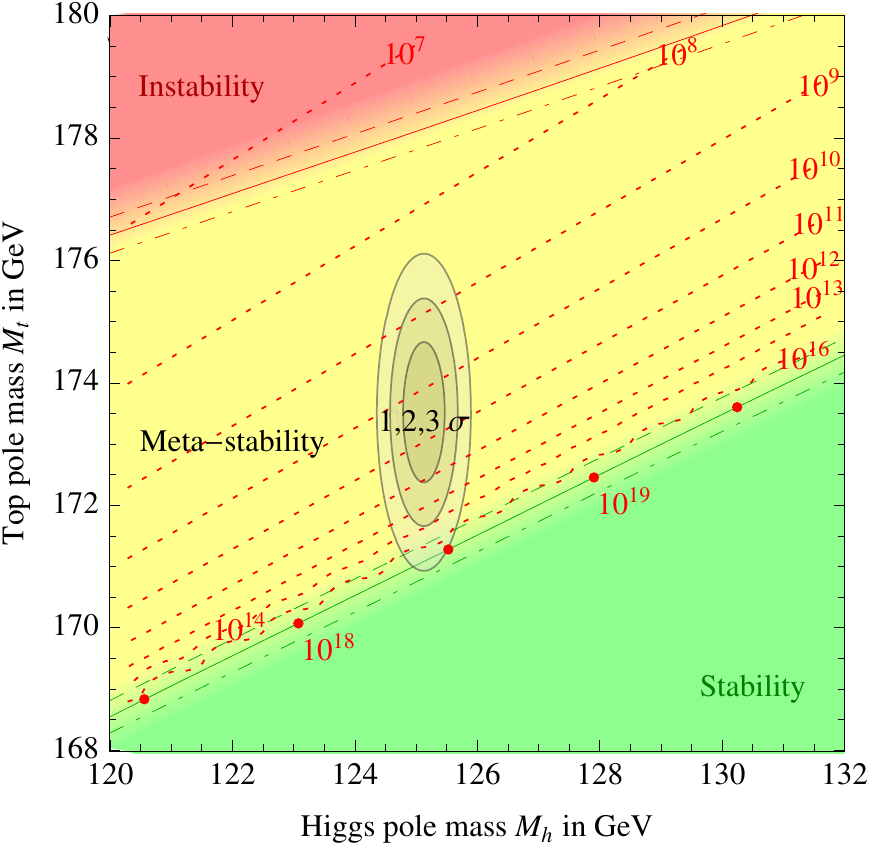} \qquad
\includegraphics[width=0.45\textwidth,height=0.45\textwidth]{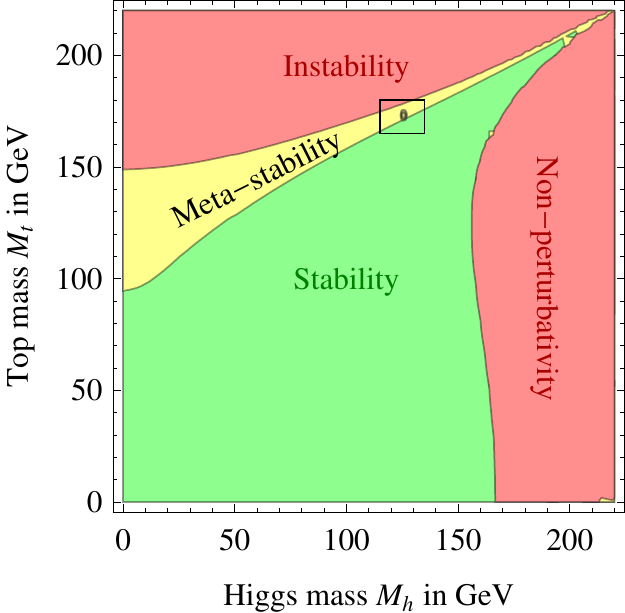}
$$
\begin{center}
\caption{\label{fig:edgy}\emph{Regions in the $\{M_h,M_t\}$ parameter space corresponding to: absolute stability (green), metastability with vacuum lifetime $\tau_{EW}>\tau_U$ (yellow), and
instability with $\tau_{EW}<\tau_U$ (red), all calculated at NNLO precision. 
The ellipses give the experimental measurements at 1, 2 and 3 $\sigma$. Red-dashed lines in the zoomed-in version on the left (from Ref.~\cite{Buttazzo}) give the instability scale in GeV. 
The zoomed-out version  in the right (from Ref.~\cite{us}) includes the region with the Higgs quartic coupling becoming non-perturbative below $M_P$. }}
 \end{center}
\end{figure}

Figure~\ref{fig:edgy} shows how the parameter space $\{M_h,M_t\}$
is split in different regions according to the large-field structure of the Higgs potential. In the green region the potential is stable, with $\lambda(\mu)>0$ for all $\mu<M_P$. In the yellow and red regions
$\lambda(\mu)$ gets negative below $M_P$ and the potential develops an instability at large field values. The difference between red and yellow regions being that in the red region the lifetime of the metastable EW vacuum is shorter than $\tau_U$. 
With the current precision of the measurements of $M_h$ and $M_t$  (see experimental ellipses) and of the theoretical calculation of the bound for stability, one concludes that  the EW vacuum is most likely metastable (given the assumptions about the absence of BSM physics). 

More precisely, the combination of $M_t$ measurements from Tevatron and LHC gives~\cite{mtcomb},
$M_t=173.34 \pm 0.76\ (0.36_{stat} \pm 0.67_{syst})$\,GeV. \footnote{Note that this world combination is already superseded by the CMS one:
$M_t=172.44 \pm 0.48\ (0.13_{stat} \pm 0.47_{syst})$\,GeV, \cite{CMSMt} but, in absence of a more up-to-date world combination, we still resort to the last one.}
While the condition for absolute stability reads~\cite{Buttazzo}:
\begin{equation}
M_t < (171.36\pm  0.15 \pm 0.25_{\alpha_s} \pm 0.17_{M_h})\,\mathrm{GeV} = (171.36 \pm 0.46)\,\mathrm{GeV}\ ,
\end{equation}
where, in the last formula, the theory error was combined in quadrature with the experimental uncertainties from
$\alpha_s(M_z)=0.1184\pm 0.0007$
\cite{alphas} and $M_h$. The theory error is an estimate of contributions from beyond-NNLO higher orders. Such small error was achieved only recently, with Refs.~\cite{shap,us,Buttazzo} being the main contributors in reaching this goal.

We see that having an EW vacuum absolutely stable up to $M_P$ requires values of $M_t$ in $\sim 2-3\sigma$ tension with the central experimental value. There is some controversy regarding the relation between the top mass that is measured at the Tevatron and LHC and the top pole-mass. Although the naive expectation would assign an error of order $\Lambda_{QCD}$ to the connection between these two numbers (or even smaller according to some educated guesses), clearly,  a better understanding of the theoretical errors
in the top mass determination would be most welcome \cite{mt}.


The right plot in figure~\ref{fig:edgy} shows the same parameter regions of the left plot [plus the "Non-perturbativity" region in which $\lambda(\mu)>4\pi$  below $M_P$] in a zoomed-out range for Higgs and top masses. This plot emphasizes that we might be living in a very special region of parameter space, really close to the critical boundary for absolute stability, in the narrow yellow wedge that corresponds to a sufficiently long-lived EW vacuum. 
This intriguing fact has motivated many speculations concerning its possible deep meaning~\cite{shap,us,Buttazzo} including: high-scale Supersymmetry~\cite{HighSUSY}, enforcing $\lambda(\Lambda)=0$ through $\tan\beta=1$; IR fixed points of asymptotically safe gravity~\cite{IRgrav},
among other ideas (even some predating the Higgs discovery~\cite{MPP}). Is $\lambda(M_P)\simeq 0$ related to the fact that we live very close to a different phase boundary, the one that separates the EW broken and unbroken phases? This second near-criticality is associated to the fact that the mass parameter in the Higgs potential, $m^2$, is extremely small in Planck units: $m^2/M_P^2\ll 1$. In relation to this, it looks as if the 
Higgs potential has a very special form at the Planck scale, with both $\lambda$ and $m^2$ being very small in natural units (not to mention the smallness of the cosmological constant). Moreover, also $\beta_\lambda$ takes a specially small value  not far from $M_P$. Why do Higgs potential parameters take these intriguing values at the Planck scale, the scale of gravitational physics, which is completely unrelated to the breaking of the EW symmetry? So far there is no compelling theoretical explanation for this.


\section{Vacuum Instability and Physics Beyond the Standard Model}

Obviously, the intriguing near-criticality discussed in the previous section would be an accidental mirage if new physics BSM appears below $M_P$ modifying the running of $\lambda(\mu)$ significantly. In fact we expect that BSM physics should be there to explain dark matter, dark energy, neutrino masses, inflation or the matter-antimatter asymmetry and it is then natural to ask how new physics could affect the near-criticality of the Higgs potential. 

There are three logical possibilities for the impact of BSM states on the stability of the Higgs potential: {\em a)} they can make the stability worse; {\em b)} they can be irrelevant; or {\em c)} they can cure it. Examples of the three options are easy to find  and I will use for illustration the simple case of type I seesaw neutrinos, which can accommodate the three cases. In seesaw scenarios, same as the top quark, neutrinos affect the running of $\lambda(\mu)$ through their Yukawa couplings, which scale as $y_\nu^2 \sim M_N m_\nu/v^2$, where $m_\nu$ is the light neutrino mass, $M_N$ is the mass of the heavy right handed neutrinos and $v=246$\,GeV is the vacuum expectation value of the Higgs field. 
The three possibilities are realized as follows: a) If $M_N$ is large enough, $y_\nu$ must be large to acommodate the same light neutrino mass and the destabilizing effect of such large Yukawas worsens the instability, potentially reducing the vacuum lifetime below $\tau_U$ [if $\lambda(\mu)<-0.05$]. This would conflict with our survival and can be used to put an upper bound on $M_N$~\cite{casas,us0} of order $M_N\simeq 10^{13\textrm{--}14}$\,GeV for $m_\nu\simeq 0\textrm{--}1$\,eV. b) If $M_N$ is smaller than the previous upper bound,  the new Yukawas are too small to alter the running of $\lambda$ significantly and their presence is irrelevant for the stability of the potential. c) A seesaw scenario that can cure the potential instability makes use of a powerful stabilization mechanism that employs a heavy singlet field $S$, with nonzero $\langle S \rangle$, coupled to the Higgs boson as $\lambda_{HS} S^2 |H|^2$ ~\cite{singlet}. Below the $S$ mass, the low-energy $\lambda$ is reduced by a threshold effect.  The apparent instability of the potential is a mirage: $\lambda$ above the $S$ threshold is really larger than the naive SM extrapolation would indicate. Such mechanism is compatible with a seesaw scenario with $M_N= \langle S \rangle$ smaller than the SM instability scale $\sim 10^{10}$\,GeV, and can also satisfy the lower bounds on $M_N$ from leptogenesis~\cite{singlet}. 

Needless to say, alternative stabilization mechanisms exist, and most extensions of the SM at the TeV scale modify the behavior (or very existence) of the Higgs at high energies/scales. In any case, potential stability argumentss (in the weak sense of demanding $\tau_{EW}\gg\tau_{U}$) can be used to constrain additional sources of instability in BSM models that generically do not guarantee (unlike Supersymmetry) a good UV behavior of the Higgs scalar  potential.

\section{Impact of physics at $M_P$ on near-criticality}

The analysis of potential stability discussed in Sects. 1 and 2 explicitly assumed that the SM is the effective theory valid below the Planck scale. The field values and energy densities involved, e.g. in a vacuum decay by tunneling, are never Planckian, so this is a consistent assumption. In particular, remember that the tunneling process is dominated by bubbles inside which the Higgs field is of the order of the scale $h_t$ where $\beta_\lambda\simeq 0$. As shown in Fig.~\ref{fig:run}, $h_t$  can be orders of magnitude larger than the instability scale but, for the values of $M_h$ and $M_t$ of experimental interest, $h_t\sim 10^{17}$ GeV, which is still sub-Planckian although not by much. 

Large Planckian effects are possible, although in the absence of a theory of quantum gravity there is no hope of calculating them. The best one can do is to try to estimate the possible impact of gravitational physics using an effective theory approach below $M_P$. In this respect there are two main possible effects one can consider. First one should include gravitational effects in the tunneling bounce, following the seminal work of Coleman and de Luccia \cite{CdL}. Generically these effects suppress decay, making the vacuum more stable (the lifetime estimates in \cite{us,Buttazzo} include such corrections). 

Another possible effect of Planckian physics is the modification of the Higgs effective potential by a tower of nonrenormalizable operators suppressed by powers of $M_P$ \cite{Branchina}. Ref.~\cite{Branchina} has analyzed the fate of the SM Higgs stability boundary after adding to the potential the terms $\lambda_6 h^6/M_P^2+\lambda_8 h^8/M_P^4$, where $\lambda_{6,8}$ are considered free parameters. For the choice $\lambda_6<0$ and $\lambda_8>0$, large changes on the stability line are found, from which very strong statements are made concerning the reliability of the near-criticality of the Higgs potential. 

Some concerns about the analysis in \cite{Branchina} are the following\footnote{Some of these issues have been discussed elsewhere, see e.g. \cite{Gino}}: 1) It does not include CdL effects in the calculation of rates, but such effects are necessarily important when the potential is affected by Planckian physics \cite{dS}. 2) It relies on an effective theory expansion in powers of $h/M_P$ that breaks down when $h$ approaches $M_P$. Simply one cannot use an effective theory close to its cutoff. 3) Toy models devised to circumvent the previous problem, e.g. by including heavy scalar and fermions degrees of freedom at the Planck scale cannot be considered seriously as representing the effects of a quantum theory of gravity. 
Besides the technical issues listed above, there is trouble with the implications derived in \cite{Branchina} from these results. Much ado is made about the sensitivity of the stability boundary in the $\{M_h,M_t\}$ plane to Planckian physics above the instability scale. This emphasis is certainly misplaced: we already showed how seesaw neutrinos heavier than the instability scale can affect dramatically that stability line  (with the difference with respect to gravitational effects that in the case of neutrinos the purely QFT calculation is under complete control). Moreover, in the case of seesaw neutrinos the origin of the BSM instability is well identified  (as due to neutrino Yukawa couplings) , while in the case of Planckian physics it is not clear why gravitational physics should make the potential more unstable (even for cases with $\lambda(\mu)>0$ for all $\mu<M_P$).

\begin{figure}[t]
$$ 
\includegraphics[width=0.6\textwidth,height=0.44\textwidth]{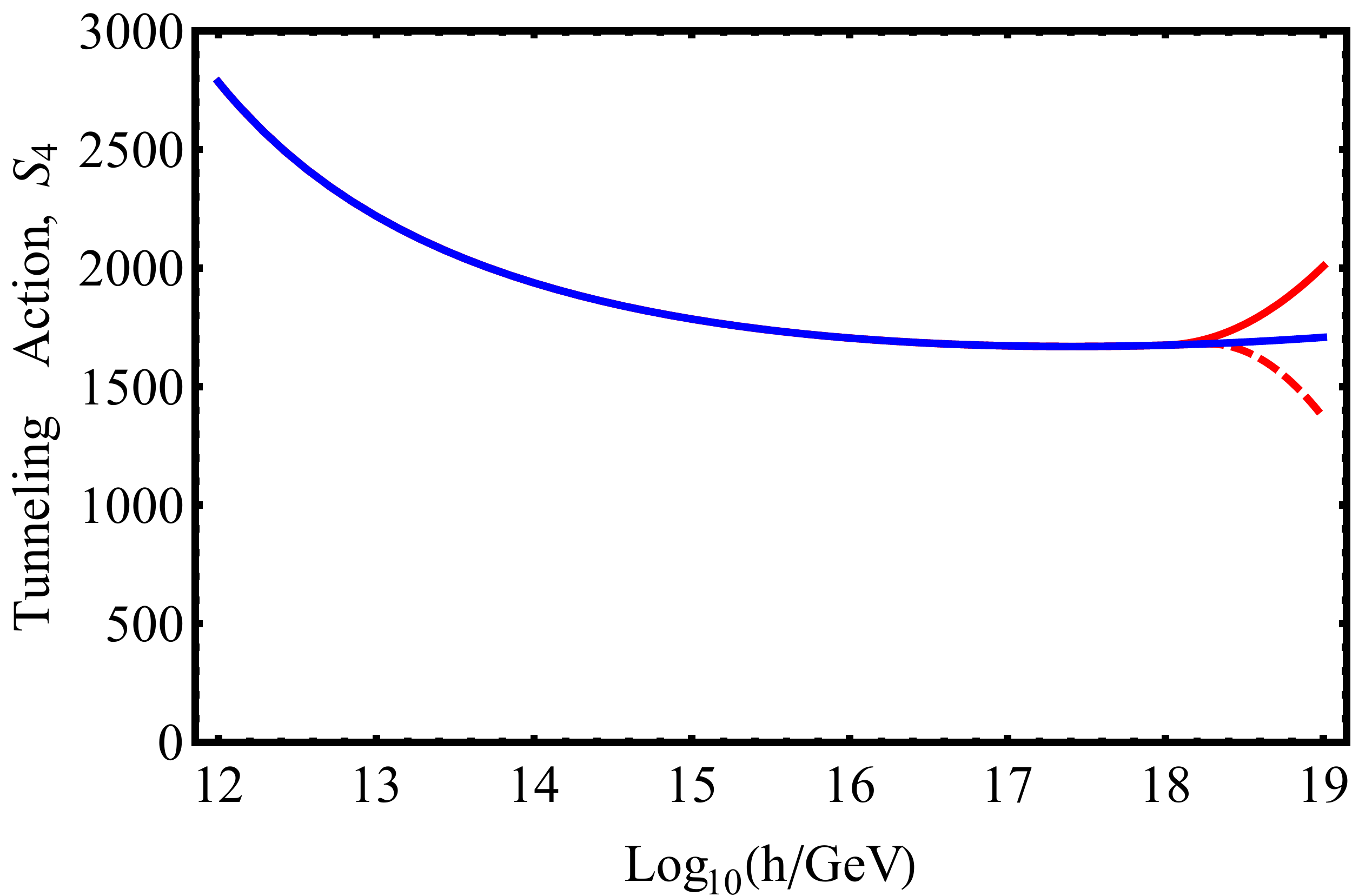} \qquad
$$
\begin{center}
\caption{\label{fig:S4}\emph{Action of the tunneling bounce as a function of the Higgs field value inside the tunneling bubble for the SM (blue) and two possible effects of Planckian physics affecting the Higgs potential and making it more stable (red) or more unstable (dashed red).}}
 \end{center}
\end{figure} 

In fact, if the gravitational corrections to the potential tend to make it more stable, then the estimate of the decay rate made in the pure SM still holds. The reason for this can be understood from Fig.~\ref{fig:S4} which shows the tunneling-bounce action as a function of the Higgs field value inside the tunneling bubble. The blue curve corresponds to the SM case, with a very flat action.
Tunneling will be dominated by the minimum value of that action, at $h\simeq h_t$. If the potential is made more unstable
by Planckian effects, one can lower the tunneling action at Higgs field values of order the Planck scale, as indicated by the dashed red line. If, on the other hand, Planckian physics make the potential more stable, the tunneling action is higher at Planckian scales but the tunneling is still dominated by the SM bounce with a negligible impact of Planckian physics on $h_t$ and the vacuum lifetime. It is in this sense that one expects (sensible) Planckian physics to decouple in the tunneling calculation. in this respect, it is interesting to notice that Planckian physics cannot be invoked to reduce significantly the gap between the stability line and the experimental values of $M_t$ and $M_h$.

\section{Conclusions}

Already from the example of seesaw neutrinos we learned that it is easier to destroy near-criticality than to explain it. However, the interest of the near-criticality of the Higgs potential hinted at by LHC is that it might be trying to tell us something deep about nature. In this respect one can compare it with gauge coupling unification. LEP-II gave us a tantalizing hint for gauge coupling unification (with a supersymmetric spectrum). LHC has given us a tantalizing hint about a possible deep reason for the near criticallity of the Higgs potential. Although admitedly grand unification rested on a more respectable theoretical foundation, it is worth considering seriously the possible theoretical reasons that might be lying behind the LHC hints of a special nature of the Higgs potential.
From this point of view, the stability of the Higgs potential is certainly a good motivation to improve (both in the experimental and theoretical fronts) the determination of the top mass,  which is the main parameter that controls how close we are to the stability line.

\section*{Acknowledgments}

This work has been partly supported by the Spanish Ministry MINECO under grants FPA2011-25948, FPA2013-44773-P and FPA2014-55613-P; by the Severo Ochoa excellence program of MINECO (grant SO-2012-0234); and by the Generalitat de Catalunya grant 2014-SGR-1450.

\end{document}